\documentstyle[aps,prl,multicol]{revtex}

\begin{document}
\input{epsf}
\draft
\preprint{}
\title{Mode Spectroscopy and Level Coupling in Ballistic Electron Waveguides
}
\author{G. Salis, T. Heinzel, and K. Ensslin}
\address{Solid State Physics Laboratory, ETH Z\"{u}rich, 8093
Z\"{u}rich,  Switzerland\\
}
\author{O. J. Homan and W. B\"achtold}
\address{Institut f\"ur Feldtheorie und H\"ochstfrequenztechnik, ETH Z\"{u}rich, 8092
Z\"{u}rich,  Switzerland\\
}
\author{K. Maranowski and A. C. Gossard}
\address{Materials Department, University of California, Santa
Barbara, Ca 93106, USA\\
}
\date{\today}
\maketitle
\begin{abstract}
A tunable quantum point contact with modes occupied in both 
transverse directions is studied by magnetotransport
experiments. We use conductance quantization of the one-dimensional 
subbands as a tool to determine the mode spectrum. A magnetic field 
applied along the direction of the current flow couples the modes. 
This can be described by an extension of the
Darwin-Fock model. Anticrossings are observed as a function of the 
magnetic field, but not for zero field or perpendicular field 
directions, indicating coupling of the subbands due to 
nonparabolicity 
in the electrical confinement.
\end{abstract}

\begin{multicols} {2}
\narrowtext

Conductance quantization in quasi one-dimensional (1D) systems 
\cite{vanWees88,Wharam88} is one of 
the crucial discoveries in the physics of semiconductor nanostructures \cite{Beenakker92}. 
Usually, such 1D channels (quantum point 
contacts - QPCs) are realized by split gate electrodes 
fabricated on Ga[Al]As heterostructures. In analogy to optical 
waveguides~\cite{Yariv}, QPCs are also known as ``ballistic electron 
waveguides''. Meanwhile, QPCs have become a key device for 
transport experiments in low-dimensional systems \onlinecite 
{Beenakker92}. Typically, they are realized in an extreme
limit where the confinement of the two-dimensional electron gas at 
the heterointerface is so strong that only the lowest two-dimensional 
(2D)
subband lies below the Fermi energy.  The mode spectrum of a QPC 
defined in such a heterostructure is dominated by the lateral confinement which can be tuned by 
appropriate voltages on the split gate electrodes. The modes, characterized 
by a single quantum number, are well separated in energy and do 
not couple to each other.

Here, we present experimental results from a ballistic electron 
waveguide realized by a split gate electrode on top of a wide 
electron system in a parabolic quantum well. The 1D energy levels can be
described by two quantum numbers for the two confining directions.
A rich mode spectrum as a function of gate voltages and magnetic fields, 
applied in different directions with respect to 
the electron waveguide, is observed.  While all levels move upwards in energy when a magnetic 
field is applied in the plane of the quantum well but perpendicular to the waveguide, 
one finds also levels moving downwards in 
energy as a function of a parallel magnetic field. The confining potential landscape inside the constriction can be studied 
by analyzing these measurements. We explain our data in terms of a generalization
of the Darwin-Fock model, which describes the energy spectrum of a
circular disc in magnetic fields perpendicular to it \cite{Darwin31}. Thus, our experiment 
is closely related to those on quantum dots \cite{Kouwenhoven97}.

Level crossings and anticrossings are observed, manifesting themselves as suppressed 
conductance plateaus. We find that non-parabolicity in the confining potential 
induces anticrossings as a function of the parallel field. Furthermore, the mode 
crossings give an absolute measure for the energy spectrum, independent of the 
lever arms with respect to gate voltages.

The parabolic quantum well consists of a $76$\,nm wide Al$_{x}$Ga$_{1-x}$As layer,
where the averaged Al content $x$ has been
varied parabolically during molecular beam epitaxy \cite{digital}. In the center of the well, 
three monolayers of Al$_{0.05}$Ga$_{0.95}$As
have been grown, serving as a probe for the position of 
the wave function in the well~\cite{Salis97,Salis99}. A $n^{+}$ doped 
layer, used as a back gate electrode, has been integrated
$1.35\,\mu$m below the well and contacted independently.
A 50\,$\mu$m wide Hall bar structure
has been defined by wet chemical etching, and the electron gas is 
accessed via Ni-AuGe Ohmic contacts. The split gate has been 
fabricated by electron beam lithography, and defines a nominally 
$400$ nm wide and $300$ nm long channel. The parabolic potential along the growth direction $z$ 
defines the waveguide in one dimension. 
The other confining direction ($y$) is
constricted by applying appropriate voltages $U_{\rm sg}$ to the split-gate 
electrode with respect to the
electron gas. The back gate can be used to change the
electron density and thus the constriction width. With $U_{\rm sg}$, one tunes the difference between the
Fermi energy $E_{\rm F}$ and the conduction band edge and thus the number of
occupied 1D subbands. 

The measurements have been performed in the mixing chamber of a $^{3}$He/$^{4}$He 
dilution refrigerator with a base temperature of 60\,mK; we estimate the electron
temperature to be about 100\,mK. At zero back-gate voltage $U_{\rm bg}$, two subbands are occupied
with sheet densities of $n_{1}= 2.6\times10^{15}$\,m$^{-2}$ and $n_{2}=2.0\times10^{15}$
\,m$^{-2}$. A  series
resistance of about $450$ $\Omega$ due to the leads between the voltage probes and the QPC is 
subtracted.

% Figure 1
\begin{figure}
\centerline{\epsfxsize=3.3 in \epsfbox{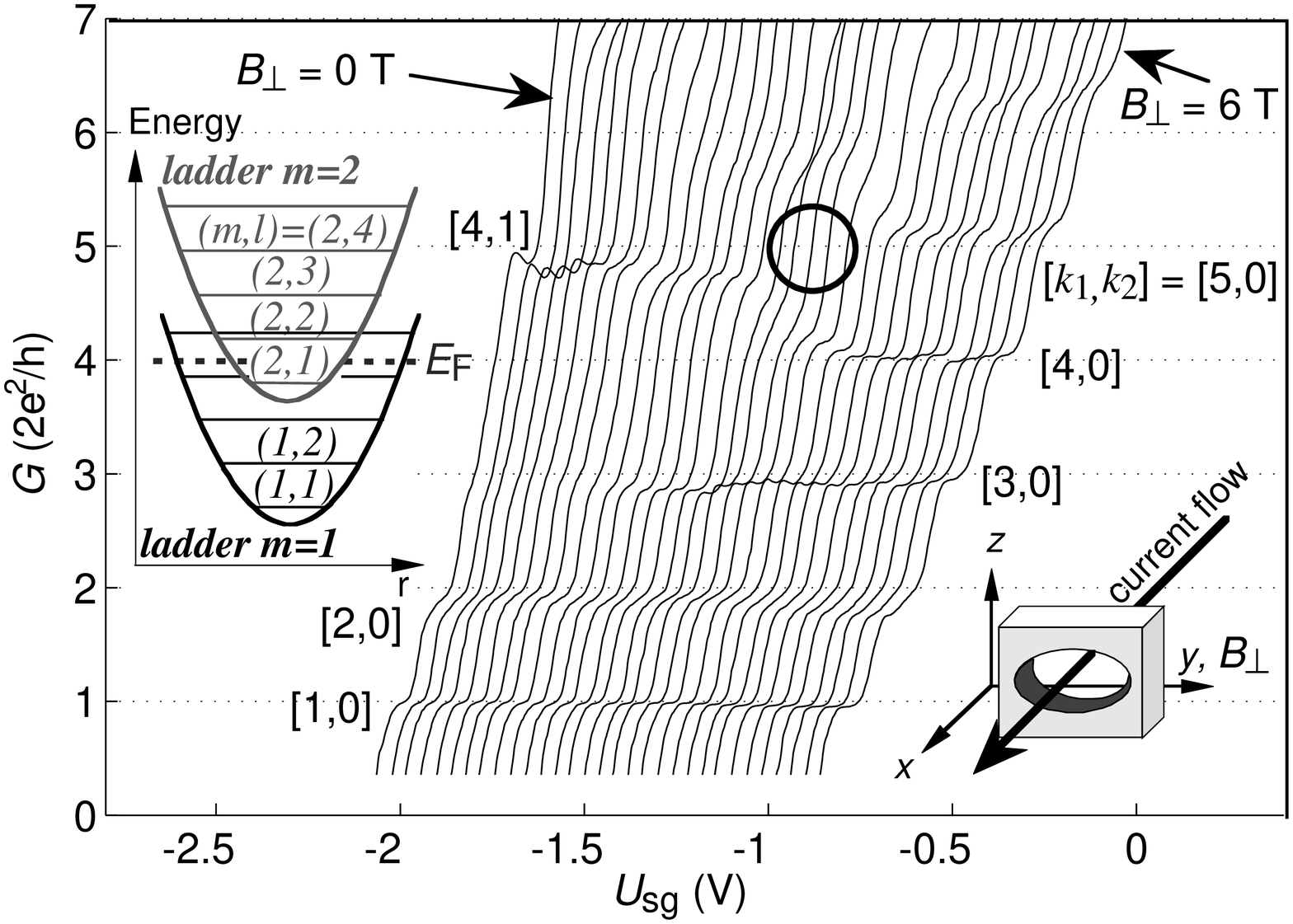}}
\caption{Measured conductance $G$ of the electron waveguide, sketched in 
the right inset, as a function of $U_{\rm sg}$ 
($U_{\rm bg}=0$) and for different $B_{\perp}$, varying between 0 and 6 T in steps of 0.2 T.
Subsequent curves are horizontally offset for clarity by 40\,mV. The 
subbands are labelled by $(m,l)$, and the number of occupied 
subbands is denoted by $[\it{k}_{1}, \it{k}_{2}]$ as explained in the text. 
Some plateaus can be suppressed and recovered by $B_{\perp}$. Left inset: scheme of the two 
subband ladders, plotted vs. spatial coordinate $r$.  In this example, 
$[\it{k}_{1}, \it{k}_{2}]$ = [4,1] is shown. The circle in the main figure denotes a situation where the 
$(1,4)$ - level is degenerate with the $(2,1)$ - level. }
\label{Fig1}
\end{figure}

We have investigated the electron waveguide in magnetic fields up to 
6\,T, in directions parallel ($B_{\parallel}$) 
and perpendicular ($B_{\perp}$) to the flow of current through the QPC
(but always in the plane defined by the quantum well).
Figure~\ref{Fig1} shows the measured conductance $G$ as a function of
$U_{\rm sg}$ for a set of $B_{\perp}$ between 0 and 
6\,T. conductance plateaus, quantized in units of $2Ne^{2}/h$, with 
$N$ being an integer, are visible up to $N=6$. While the lowest two plateaus are only 
slightly modified by $B_{\perp}$, the higher
plateaus vanish for certain ranges of $B_{\perp}$. The circle
in Fig.~\ref{Fig1} marks a regime in which the $N=5$ plateau is suppressed. Similar
suppressions are observed when $B_{\parallel}$ or $U_{\rm bg}$ is
varied (not shown).

The energy levels of our electron waveguide are labelled 
by two quantum numbers $l$ and $m$, belonging to the two confinement directions in 
$y$- and $z$-directions, respectively. A ladder of 
1D states is attributed to each of the two occupied 2D subbands. Our system can thus be seen as being composed of $2$ coupled 
QPCs in parallel \cite{Salis2}. In this respect, it is similar to two separated QPCs arranged in parallel, 
as investigated in Refs.~\onlinecite {Smith89,Simpson93,Simmons97,Thomas99}. However, the
1D ladders in our sample reside in identical host potentials and have large wave 
function overlaps. We label these states by $(m,l)$, where $m$=$\{1,2\}$ denotes the 
ladder to which the state belongs, and $l$ labels the states inside 
ladder $m$. Furthermore, the number of occupied subbands 
attributed to ladder $m$ is denoted by by $k_{m}$. The total number $N$ of 1D levels below the Fermi energy 
$E_{F}$ is thus given by $N=k_{1}+k_{2}$.  
The suppression of the plateau at $N=5$, for example, can be understood as a 
degeneracy of the levels $(1,4)$  and $(2,1)$. If such two 
degenerated levels cross the Fermi energy, the 
conductance changes by $4e^2/h$. Therefore, no 
plateau at $N=5$ is observable in this example.

\begin{figure}
\centerline{\epsfxsize=3.5 in \epsfbox{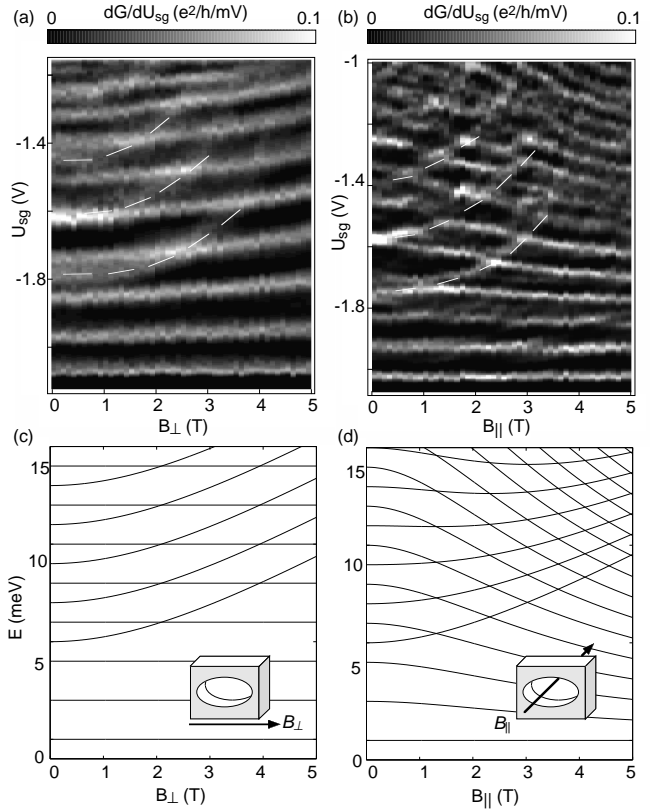}}
\caption{Grayscale plot of $dG/dU_{\rm sg}$ as a
function of $U_{\rm sg}$ and $B_{\perp}$ (a),  and $B_{\parallel}$, 
respectively (b).
A high transconductance is related to a crossing between $E_{\rm F}$ 
and a 1D subband level. The dashed lines are guides to the eye. In (c) and (d), 
calculated energy levels for a parabolically confined waveguide 
with $\omega_{y}=2$\,meV and $\omega_{z}=5$\,meV are shown.}
\label{Fig2}
\end{figure}

The energy spectrum and its dependence on magnetic field can be visualized by 
plotting the transconductance $dG/dU_{\rm sg}$ with respect to $U_{\rm 
sg}$ and magnetic field [Fig.~\ref{Fig2}(a) and (b)]. 

% Figure 2

On a plateau in $G(U_{\rm sg})$, $E_{\rm F}$ lies between two
1D levels. This corresponds to a low transconductance (dark 
regions). On the other hand, on a transconductance peak a 1D
level crosses $E_{\rm F}$ (bright regions). With increasing 
$U_{\rm sg}$, the subband ladders move upwards with respect to 
$E_{\rm F}$ and thus the mode spectrum can be scanned. The bright lines in
Fig.~\ref{Fig2} reflect the 1D levels plotted vs. $B$.  
In both cases, two ladders of 1D subbands are observed
whose magnetic field behavior depends on both the ladder index $m$ as
well as on the direction of the magnetic field. 
In perpendicular fields, nine levels of ladder 1 are
visible, showing a weak shift towards higher energies in increasing
magnetic fields, while the energy separation between
subsequent levels increases. The four visible modes
belonging to the second ladder show a much stronger 
upwards shift in energy with increasing magnetic field. In
parallel magnetic fields, the mode spectrum shows two striking 
differences. First of all, the states belonging to ladder 1 shift downwards in energy for 
$B_{\parallel}>1$\,T, while the states of the second ladder are 
essentially insensitive to the direction of the magnetic field. Second, 
anticrossings form between 
$m$ = $1$ - states and $m$ = $2$ - states (Fig. 3).

By varying $U_{\rm sg}$, not only the subband ladders are displaced 
with respect to $E_{\rm F}$,  but also the step-size of the ladders changes due to the modification 
of the waveguide shape. In addition, the lever arm $\alpha = dE/dU_{\rm 
sg}$ may not be perfectly constant. Therefore, $U_{\rm sg}$ can only be
a coarse measure for the energy. We can estimate $\alpha$ by measuring the
smearing-out of a conductance plateau as a function of the source-drain
voltage~\cite{Kouwenhoven89}, and find an average subband-spacing of
2\,meV at $B=0$, which has to be compared with a step width in $U_{\rm sg}$. We obtain 
$\alpha \approx 0.02meV/mV$. However, the energy scale is fixed 
by the crossing points of mode levels.
In the following, we compare the measured mode spectrum with a
calculation of the energy levels of a 2D constriction in a magnetic 
field. For the $B_{\perp}$-case [Figs. 2(a) and (c)], the magnetic field 
along the $y$-direction modifies the dispersion relation in
$x$-direction and thus the effective mass,
whereas the $z$-confinement is enhanced due to the diamagnetic
term proportional to $B_{\perp}^2$. A coupling term of the $z$- and
$x$-direction results in a shift of the Fermi surface in 
$k_{x}$-direction. Neglecting electron-electron interactions,
no coupling of the $y$- and $z$
direction is expected for a separable confinement potential $V(y,z)$. For a parabolic confinement
$V(y,z)=m^*(\omega_{y}^2y^2+\omega_{z}^2z^2)/2$ the subband energies
are given by
$E_{\it{ml}}=\hbar\omega_{y}(\it{l}-\frac{1}{2})+\hbar\sqrt{\omega_{z}^2+\omega_{\perp}^2}(m-
\frac{1}{2})$, where $\omega_{\perp}$ is the cyclotron frequency $eB_{\perp}/m^{*}$ \cite{Scherbakov96}.
Figure~\ref{Fig2}(c) shows the energy fan for $\omega_{y}=2$\,meV and
$\omega_{z}=5$\,meV. The levels cross each other at identical 
magnetic fields and without level repulsions. 
The measured spread of the step size with increasing $B_{\perp}$ [Fig.~\ref{Fig2}(a)] can be
explained by a field-dependent reduction of the density of states in 
the 2DEG, due to the modified dispersion relation, leading to a decrease of $E_{\rm F}$
relative to the conduction band bottom with increasing $B_{\perp}$ \cite{Stern68,Salis98}.

$B_{\parallel}$ couples the two confining directions $y$ and
$z$. If a parabolic confinement with rotational symmetry 
is assumed, this leads to the ordinary
Zeeman effect, described by the Darwin-Fock states~\cite{Darwin31}.
Within this model, the eigenstates of a circular disc (or a wire with
a circular cross section, respectively)
with different angular momenta are
degenerate at $B_{\parallel}=0$. This degeneracy is lifted by non-zero
$B_{\parallel}$, and, depending on its angular momentum, the energy of
one level can either increase or decrease. For a non-circular confinement potential
as realized in our QPC, the degeneracy is
already lifted at $B=0$.
% Figure 3

\begin{figure}
\centerline{\epsfxsize=3.2 in \epsfbox{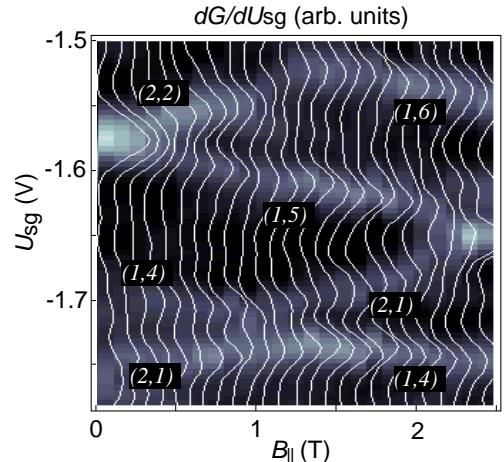}}
\caption{Close-up of Fig.~\ref{Fig2}(b) showing anticrossing of the
 $(2,1)$ - level with the $(1,4)$ - level. In addition to the grayscale,
the individual transconductance traces are shown.}
\label{Fig3}
\end{figure}

A calculation for parabolic confinements with arbitrary $\omega_{y}$ 
and $\omega_{z}$ 
gives the result~\cite{Schuh85}

\begin{equation}
E_{\it{ml}}=\hbar\omega_1(\it{l}-\frac{1}{2})+\hbar\omega_2(m-\frac{1}{2}),
\label{equation1}
\end{equation}

with

\begin{equation}
\omega_{1,2}^2=\frac{1}{2}(\omega_{\parallel}^2 +
\omega_{y}^2+\omega_{z}^2 \pm \sqrt{ \left
(\omega_{\parallel}^2+\omega_{y}^2+\omega_{z}^2\right)^2  - 
\omega_{y}^2
\omega_{z}}.
\end{equation}

Here we used
$\omega_{\parallel}=eB_{\parallel}/m^*$. 
The larger of the two frequencies $\omega_{1}$
and $\omega_{2}$ increases with $B_{\parallel}$, while
the smaller one decreases. Figure~\ref{Fig2}(d) shows a
calculation for $\omega_{y}=2$\,meV and $\omega_{z}=5$\,meV, which
reproduces the observed decrease of the 1D subband levels of
Fig.~\ref{Fig2}(b). There are two qualitative differences between the 
simple model of Eq.~\ref{equation1} and the measurements: First, the 
levels do not cross at the same $B_{\parallel}$ as in the model, and 
second the measured levels show anticrossings.
The anticrossing is particularly pronounced between, for example, 
$(1,4)$ and $(2,1)$, while it is much weaker for other levels, 
like $(1,5)$ and $(2,1)$ (Fig.~\ref{Fig3}).

As seen in Fig.~\ref{Fig2}(a), no anticrossing is observed for 
$B_{\perp}$. Similarly, the traces of transconductance maxima
cross regularly as a function of 
$U_{\rm bg}$ for $B=0$. This indicates that the electric confinement potential 
is separable in $y$ and $z$-directions. However, for finite fields 
$B_{\parallel}$, the magnetic confinement prevents a decoupling of the Hamiltonian into
two independent oscillators if the QPC confinement is not exactly
parabolic, leading to level anticrossings. We therefore argue that 
the observed anticrossings have their origin in nonparabolicities
of the confining potential. For square well electric confinements, a 
complicated level structure with unregular anticrossings was 
calculated~\cite{Hansen89,Robnik84}. In our samples, the 
dominant non-parabolicity may come from
electron-screening, giving rise to a more rectangularly shaped 
effective potential in $z$-direction. We have calculated the 
influence of the three monolayers wide potential perturbation in the center
of the quantum well, and obtain  anticrossings in first-order perturbation theory, 
which are too small to account for the observed effects.\\
No clear signature of level locking could be observed, although
predicted for coupled one-dimensional wires~\cite{Sun94}.

For the Darwin-Fock model, it is irrelevant whether the waveguide
extends to infinity in the direction parallel to the magnetic
field, the system is a truly 2D disc. Quantum dots are usually described 
as such 2D discs and have not only been thoroughly studied
theoretically, but also widely investigated by magnetotransport
measurements~\cite{Kouwenhoven97}. In those experiments, the dot
is weakly coupled to two leads via tunnel barriers. Such small quantum
dots with sizes comparable to the cross section of our waveguide are
highly interesting also from a theoretical point of view, since
interaction effects, in particular Coulomb blockade as well as 
exchange and correlation corrections, and their influence on the energy 
spectrum for small occupation numbers can be
experimentally determined. However, it is difficult to distinguish
between effects originating from the shape of the confining potential
and those from interaction effects. In that sense, our QPC provides 
a perfect reference system, since by effectively extending the dot to infinity
in the direction of the magnetic field, interaction effects are essentially
turned off. As can be seen from the above measurements, the magnetic
field dependence of the energy levels is nontrivial even in the
absence of interactions.

In conclusion, we have experimentally investigated the 1D subband 
energies of a ballistic electron waveguide realized in a semiconductor parabolic quantum well by 
measuring its
transconductance as a function of gate voltage and magnetic fields, 
aplied in different directions.  
Depending on their mode index, energy levels may shift upwards or
downwards in energy under parallel magnetic fields, while they always
shift to higher energies in perpendicular magnetic fields. We have
modelled this behavior in terms of an analytical model.
In the case where $B$ is oriented
perpendicular to the axis of the QPC, coupling between 
one-dimensional
subbands is neither observed nor expected. For field directions along 
the axis, a coupling, manifested as level anticrossings, is
observed and interpreted as a consequence of nonparabolic
confinement.

We acknowledge valuable discussions with A. Lorke and J. P. Kotthaus. 
This project was financially supported by the Swiss
Science Foundation, AFOSR grant F 49620-94-1-0158 and the NSF Center 
for Quantized Electronic Structures (QUEST).

\end{multicols}

\begin{references}
\bibitem{vanWees88}B. J. van Wees, H. van Houten, C. W. J. Beenakker,
J. G. Williamson, L. P. Kouwenhoven, D. van der Marel, and C. T.
Foxon, Phys. Rev. Lett. {\bf 60}, 848
(1988).
\bibitem{Wharam88}D. A. Wharam, T. J. Thornton, R. Newbury, M.
Pepper, H. Ahmed, J. E. F. Frost, D. G. Hasko, D. C. Peacock, D. A.
Ritchie, and G. A. C. Jones, J. Phys. C {\bf 21}, L209 (1988).
\bibitem{Beenakker92} for a review, see H. van Houten, C.W. J. 
Beenakker, and B.J. van Wees, {\it Quantum Point Contacts}, in Semicond. and
Semimet. {\bf 35}, ch. 2 (1992).
\bibitem{Yariv}A. Yariv, {\it Quantum Electronics} 3rd ed., Wiley,
New York 1989.
\bibitem{Darwin31}C. G. Darwin, Proc. Cambridge Philos. Soc. {\bf 27},
86 (1931); V. Fock, Z. Phys. {\bf 47}, 446 (1928).
\bibitem{Kouwenhoven97}for a review, see L.P. Kouwenhoven, C.M. Marcus, P.L. McEuen, S.
Tarucha, R.M. Westervelt, and N.S. Wingreen, pp. 2-110, in Mesoscopic 
Electron Transport, Kluwer 1997.
\bibitem{digital}A. C. Gossard, IEEE J. Quant. Electron. {\bf
22}, 1649 (1986).
\bibitem{Salis97}G. Salis, B. Graf, K. Ensslin, K. Campman, K. 
Maranowski, and
A. C. Gossard, Phys. Rev. Lett. {\bf 79}, 5106 (1997).
\bibitem{Salis99}G. Salis, P. Wirth, T. Ihn, T. Heinzel, K. Ensslin,
K. Campman, K. Maranowski, and A. C. Gossard, to appear in  Phys. 
Rev. B {\bf 59},(1999).
\bibitem{Salis2}G. Salis, T. Heinzel, K. Ensslin, O.J. Homan, W. 
B\"achtold, K. Maranowski, and A.C. Gossard, Proceedings of the 23rd 
International Conference on the Physics of Semiconductors, in print.
\bibitem{Smith89}C. G. Smith, M. Pepper, R. Newbury, H. Ahmed, D. G.
Hasko, D. C. Peacock, J. E. F. Frost, D. A. Ritchie, G. A. C. Jones,
and G. Hills, J. Phys.: Condens. Matter  {\bf 1}, 6763 (1989).
\bibitem{Simpson93}P. J. Simpson, D. R. Mace, C. J. B. Ford, I.
Zailer, M. Pepper, D. A. Ritchie, J. E. F. Frost, M. P. Grimshaw, and
G. A. C. Jones, Appl. Phys. Lett. {\bf 63}, 3191 (1993).
\bibitem{Simmons97}I. M. Castleton, A. G. Davies, A. R. Hamilton, J.
E. F. Frost, M. Y. Simmons, D. A. Ritchie, and M. Pepper, Physica B 
{\bf 251}, 157 (1998).
\bibitem{Thomas99} K.J. Thomas, M. Y. Simmons, W.R. Tribe, A. G. Davies,  and M. Pepper, 
cond-mat/9901161 (1999).
\bibitem{Kouwenhoven89}L. P. Kouwenhoven, B. J. Wees, C. J. P. M.
Harmans, J. G. Williamson, H. van Houten, C. W. J. Beenakker, C. T.
Foxon, and J. J. Harris, Phys. Rev. B {\bf 39}, 8040 (1989).
\bibitem{Scherbakov96}A. G. Scherbakov, E. N. Bogachek and U.
Landman, Phys. Rev. B {\bf 53}, 4054 (1996).
\bibitem{Stern68}F. Stern, Phys. Rev. Lett. {\bf 21}, 1687 (1968).
\bibitem{Salis98}G. Salis, B. Ruhstaller, K. Ensslin, K. Campman, K.
Maranowski,
and A. C. Gossard, Phys. Rev. B {\bf 58}, 1436 (1998).
\bibitem{Schuh85}B. Schuh, J. Phys. A {\bf 18}, 803 (1985).
\bibitem{Hansen89}W. Hansen, T. P. Smith III, K. Y. Lee, J. A. Brum,
C. M. Knoedler, J. M. Hong, and D. P. Kern, Phys. Rev. Lett. {\bf 62},
2168 (1989).
\bibitem{Robnik84}M. Robnik, J. Phys. A {\bf 19}, 3619 (1984).
\bibitem {Sun94}Y. Sun and G. Kirczenow, Phys. Rev. Lett. {\bf 72}, 
2450 (1994).
\end{references}
\end{document}